\documentclass[prd,showpacs,floatfix,amsmath,amssymb,nofootinbib,twocolumn]{revtex4-2}
\usepackage{graphicx,color,dcolumn,booktabs,bm}
\usepackage{longtable,comment,braket}
\usepackage{times}
\usepackage{overpic}
\usepackage{amssymb,tipa}
\usepackage{indentfirst}
\usepackage{feynmf}   
\usepackage{slashed}  
\usepackage{cases}
\usepackage{psfrag}
\usepackage{subfigure}
\usepackage{color}
\usepackage{multirow}
\usepackage{lineno}
\usepackage[dvipsnames]{xcolor}

\usepackage[colorlinks, citecolor=blue,anchorcolor=red,menucolor=red,linkcolor=blue,filecolor=red,runcolor=red,urlcolor=blue,frenchlinks=red]{hyperref}
\usepackage{cleveref}

\usepackage{ulem}
\def\be{\begin{equation}}
\def\ee{\end{equation}}

\begin{document}
\title{Chiral phase transition and spin alignment of vector meson in the Polarized-Polyakov-loop Nambu--Jona-Lasinio model under rotation}
\author{Fei Sun$^{a,b,c}$}
\email{sunfei@ctgu.edu.cn}
\author{Jingdong Shao$^{b,d,e}$}
\email{shaojingdong19@mails.ucas.ac.cn}
\author{Rui Wen$^{b}$}
\email{rwen@ucas.ac.cn}
\author{Kun Xu$^{b}$}
\email{xukun21@ucas.ac.cn}
\author{Mei Huang$^{b}$}
\email{huangmei@ucas.ac.cn}
\affiliation{$^a$Department of Physics, China Three Gorges University, Yichang, 443002, China \\
$^b$School of Nuclear Science and Technology, University of Chinese Academy of Sciences, Beijing 100049, China\\
$^c$Center for Astronomy and Space Sciences, China Three Gorges University, Yichang 443002, China\\
$d$ School of Physical Sciences, University of Chinese Academy of Sciences, Beijing 100049, China\\
$e$ Institue of High Energy Physics, Chinese Academy of Sciences,Beijing, 100049, China}

\begin{abstract}

By using the extrapolation method, a polarized Polykov-loop potential at finite real angular velocity is constructed from the lattice results at finite imaginary angular velocity. The chiral and deconfinement phase transitions under rotation have been simultaneously investigated in the Polarized-Polyakov-loop Nambu-Jona-Lasinio (PPNJL) model. It is observed that both critical temperatures of deconfinement and chiral phase transition increase with the angular velocity, which is in consistent with lattice results. The spin alignment of vector meson has the negative deviation of $\rho_{00} -1/3$ under rotation, and the deviation in the PPNJL model is much more significant than that in the NJL model and the quark coalescence model, which revealing the important role of rotating gluons on the quark polarization.

\end{abstract}

\maketitle

\section{Introduction\label{sec1}}
 
For nearly a century, it has been recognized that the coupling between spin and orbital angular momentum plays a crucial role, for example in the Barnett effect\cite{A.Einstein,S.J.Barnett,Barnett:1935wyv}. This effect, which manifests as an increase in magnetization due to rotation, is influenced by the intricate interplay between these two types of angular momentum. In recent two decades, spin-orbit interactions in non-central heavy-ion collisions has attracted lots of interests. 

In non-central ultra relativistic heavy ion collisions, the angular momentum with the magnitude of  $\mathcal{O}(\text{10}^4-\text{10}^5) \hbar $ \cite{Becattini:2007sr} with local angular velocity in the range of $0.01\sim 0.1$ GeV and the angular velocity of order $\omega \sim 10^{21}\text{s}^{-1}$ \cite{Li:2017slc} can be created. The spin-orbit coupling of quantum chromodynamics (QCD) could give rise to numerous intriguing phenomena. For example,  Liang and Wang in 2005 predicted the global spin polarization for the final-state hyperons \cite{Liang:2004ph} as well as vector meson spin alignment \cite{Liang:2004xn}. The STAR Collaboration observed the global polarization of $\Lambda$ and $\bar{\Lambda}$ hyperons in non-central heavy-ion collisions in 2017 \cite{STAR:2017ckg,STAR:2018gyt,STAR:2020xbm}. In 2020, the ALICE Collaboration reported spin alignment of $\phi$ and $K^{*0}$, and $\rho_{00}<1/3$ for both $\phi$ and $K^{*0}$ vector mesons with small transverse momenta \cite{ALICE:2019aid}. The STAR Collaboration also reported the measurement of spin alignment for $\phi$ and $K^{*0}$ vector mesons, with the global spin alignment for $\phi$ yielding unexpected results that $\rho_{00}>1/3$ with a large deviation from $1/3$, while $\rho_{00}\approx 1/3$ for $K^{*0}$ \cite{STAR:2022fan} in 2023. The reasons for the varied results in the spin alignments of the two vector mesons remain unclear, prompting a series of extensive studies \cite{Sheng:2019kmk,PhysRevD.105.099903, Sheng:2020ghv,Xia:2020tyd,Muller:2021hpe, Yang:2021fea,Goncalves:2021ziy,Li:2022neh, Wagner:2022gza,Wagner:2022gza,Kumar:2022ylt,Sheng:2022wsy,Sheng:2022ffb,Kumar:2023ghs,Kumar:2023ojl}.

Meanwhile, QCD matter under rotation has also attracted much attention in recent years.
The chiral symmetry breaking and its restoration, and confinement-deconfinement phase transition are most important topics in QCD. In non-central heavy-ion collisions, the system carries large magnitude of orbital angular momentum to create a state of rotating Quark-Gluon Plasma (QGP). This raises the natural question whether rotation affects the QCD phase transition and how the transition  evolves with rotation. Currently, there has been considerable theoretical research dedicated to this subject. Specifically, in relation to the chiral phase transition, there have been numerous comprehensive investigations in Refs. \cite{Chen:2015hfc,Jiang:2016wvv,Ebihara:2016fwa, Chernodub:2016kxh,Chernodub:2017ref,Wang:2018sur,Sun:2021hxo, Xu:2022hql,Sun:2023kuu}. These studies have revealed that rotation suppresses the chiral critical temperature. In contrast, the impact of rotation on the deconfinement phase transition is less understood. Using the holographic QCD approach, it has been found in Ref. \cite{Chen:2020ath} that the deconfinement critical temperature decreases with increasing angular velocity, which is supported by other holography studies \cite{Chen:2023cjt,Braga:2023qej,Li:2023mpv, Ambrus:2023bid,Zhao:2022uxc,Golubtsova:2022ldm, Chen:2022mhf,Chen:2022smf,Yadav:2022qcl,Braga:2022yfe, Cartwright:2021qpp,Golubtsova:2021agl,Chernodub:2020qah}. A hadron resonance gas model obtained results are in consistent with those of holographic QCD models \cite{Fujimoto:2021xix}. However, recent lattice QCD simulations \cite{Braguta:2021jgn,Braguta:2022str,Yang:2023vsw} have suggested that the critical temperature of the confinement phase transition in gluon dynamics rises with increasing angular velocity, which is opposite to above results obtained in effective models. 

The inconsistency between the effective model calculations and lattice results on rotation 
could be due to the absence of non-perturbative gluonic effects in the model calculations. 
The quarks have the spin of $1/2$, and gluons have the spin of $1$, in principle, gluons are more sensitive to rotation due to the coupling between the spin and orbital angular momentum. It has been observed in \cite{WeiMingHua:2020eee} that the spin-1 vector meson shows Zeeman-like behavior under rotation, and the mass of the spin component $s_z = 1$ vector meson decreases linearly with $\omega$, while constituent quark mass almost does not change under small angular velocity. This indicates that spin-1 vectors are more sensitive to rotation than that of spin-$1/2$ quarks. It is reasonable to expect that spin-1 gluons exhibit similar behavior to the vector mesons under rotation.

In \cite{Chen:2015hfc,Jiang:2016wvv,Ebihara:2016fwa, Chernodub:2016kxh,Chernodub:2017ref,Wang:2018sur,Sun:2021hxo, Xu:2022hql},  the investigation of chiral phase transition under rotation didn't take into account the effect from rotating gluodynamics. Therefore, in this work, we try to investigate how rotating gluodynamics would affect the chiral phase transition. Recently, the rotating gluon system has been considered in \cite{Cao:2023olg,Jiang:2023hdr}. In this work, we will construct the polarized Polyakov-loop potential induced by rotation based on the lattice results of Ref. \cite{Braguta:2021jgn}, and add the 2-flavor NJL model on this polarized Polyakov-loop potential, we call this model as PPNJL. Then investigate the chiral phase transition of QCD and the spin alignment of the vector meson $\rho$. The PPNJL model was chosen for its effectiveness in describing chiral symmetry breaking and the confinement mechanism \cite{Meisinger:1995ih,Meisinger:2001cq,Fukushima:2003fw,Mocsy:2003qw,Megias:2004hj,Ratti:2005jh,Fukushima:2008wg}.

The organization of this work is as follows. We construct a polarized Polykov-loop potential based on lattice results in Section~\ref{sec2}. In Section~\ref{sec3}, we add 2-flavor NJL model on Polarized Polykov-loop potential, and construct the  2-flavor PPNJL model, and introduce how to calculate the quark polarization as well as the spin alignment of the vector meson. Then in Section~\ref{sec4}, we present numerical results and discuss the effective quark mass, Polyakov-loop, the chiral and deconfinement phase transitions, as well as the spin alignment of the $\rho$ meson under rotation. Finally, we give summary and discussion at the end of the paper.\\

\section{Polarized Polyakov-loop potential under rotation}
\label{sec2}

The first lattice study of rotating QCD matter was conducted in Ref. \cite{Yamamoto:2013zwa}. Recently, a series of studies on the properties of SU(3) gluodynamics under rotation using lattice simulations have been undertaken \cite{Braguta:2022str,Braguta:2023yjn, Braguta:2023kwl,Yang:2023vsw,Chernodub:2022veq,Braguta:2023qex,Braguta:2023iyx}. Due to the sign problem, the Monte Carlo simulations are conducted with the imaginary angular velocity and the deconfinement phase transition is found shifted to lower temperature as the imaginary angular velocity increases \cite{Braguta:2022str,Braguta:2023yjn, Braguta:2023kwl,Yang:2023vsw}.

\begin{figure}[t]
\setlength{\unitlength}{1mm}
\centering
\includegraphics[width=0.5\textwidth]{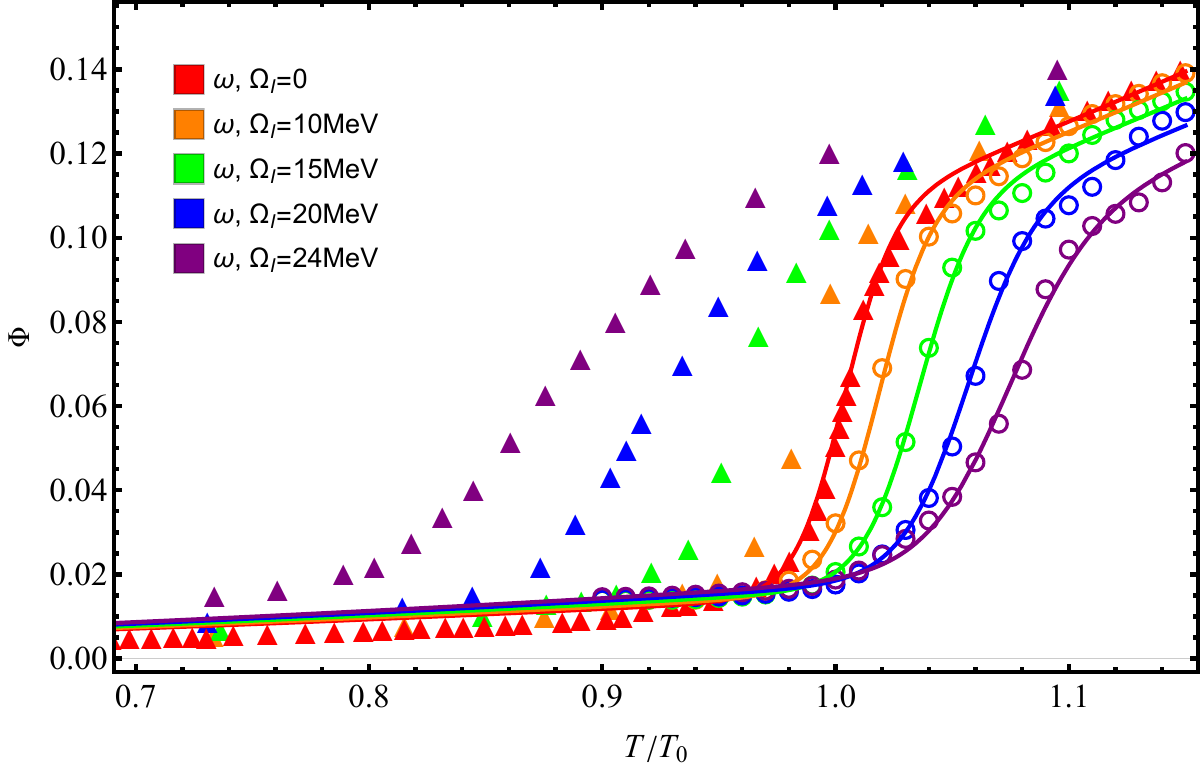}
\caption[]{(Color online) The Polyakov-loop as a function of the temperature scaled by $T_{0}$ for different values of angular velocity $\omega$. The triangle dots correspond to lattice data with respect to different imaginary angular velocity in Ref. \cite{Braguta:2021jgn}, the hollow dots correspond the analytical continuation to real angular velocity and the solid lines correspond the fitting results with Eq. (\ref{eq:phifit}). }
\label{PPloop.pdf}
\end{figure}

In order to effectively take into account the rotating gluon background, we construct the polarized Polyakov-loop potential induced by rotation based on the lattice results of Ref. \cite{Braguta:2021jgn}. To investigate the influence of rotation on gluon dynamics properties, we take the rigid cylinder rotating around the z-axis with constant angular velocity $\omega$. The rotational system must satisfy the causality condition $\omega r <1$, and thus, in principle, the boundary conditions should be taken into account. In Ref. \cite{Braguta:2021jgn}, three types of boundary conditions including open boundary conditions (OBC), periodic boundary conditions (PBC) and Dirichlet boundary conditions (DBC)) were considered. However, the results indicate that the qualitative behavior does not depend on the boundary conditions. 
In our calculations, we choose the lattice result in Ref. \cite{Braguta:2021jgn} in the case of PBC. 

In Fig. \ref{PPloop.pdf}, the triangle dots are the Polyakov-loop $\Phi$ in \cite{Braguta:2021jgn} as a function of the scaled temperature $T/T_0$ with different $\Omega_{I}$. In order to do the analytical continuation, we employ $\Omega_{I}\rightarrow i \omega$ and expand the Polyakov-loop $\Phi$ by rescaling of temperatures
\begin{align}\label{eq:phiext}
    \Phi(T,\omega)=\Phi(T',\omega=0),\quad T=T'+k_2(T')\omega^2+k_4(T')\omega^4.
\end{align}
Due to the small angular velocities, we only consider first 4-order of $\omega$. With Eq. (\ref{eq:phiext}), we extrapolate the Polyakov-loop $\Phi$ from imaginary angular velocity to real angular velocity, which are shown as hollow dots in Fig. \ref{PPloop.pdf}. We introduce an analytical function $f\left( {T,\omega } \right)$ to fit the Polyakov-loop empirically in a certain temperature interval
\begin{align}\label{eq:phifit}
    \Phi\left(T,\omega \right)=\left(\frac{T}{T_0}\right)^2f\left( {T,\omega } \right),
\end{align}
with
\begin{eqnarray}
f\left( {T,\omega } \right) = a\left( \omega  \right)\tanh \left( {b\left( \omega  \right)\left( {\frac{T}{{{T_0}}} - c\left( \omega  \right)} \right)} \right) + d\left( \omega  \right),
\label{ppotential2}
\end{eqnarray}
here $x(\omega)=x_{0}+x_{2}\omega^{2}+x_{4}\omega^{4}$ for $x\in\{a,b,c,d\}$ with $\omega$ values in unit of MeV. The corresponding coefficients are determined by setting Polyakov-loop $\Phi$ consistent with the analytical continuation results, which are listed in TABLE. \ref{table_parameters}. We plot them as solid lines in Fig. \ref{PPloop.pdf}. It is observed that indeed, though the critical temperature for deconfinement phase transition is shifted to lower temperature as the imaginary angular velocity increases, while the critical temperature is shifted to higher temperature as the angular velocity increases.
\begin{figure}
\setlength{\unitlength}{1mm}
\centering
\includegraphics[width=0.5\textwidth]{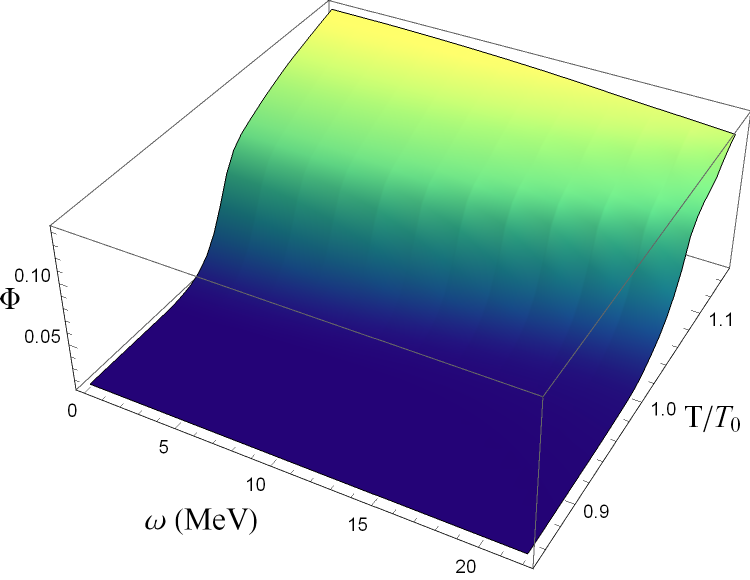}
\caption[]{(Color online) The 3D plot of the Polyakov-loop as functions of the angular velocity and the scaled temperature $T/T_0$. }
\label{Ploopr.pdf}
\end{figure}

In Fig. \ref{Ploopr.pdf}, we present a 3D plot of the Polyakov-loop $\Phi$ as a function of the real angular velocity $\omega$ and the scaled temperature $T/T_0$. It is evident from this figure that the critical temperature of the confinement-deconfinement phase transition increases with the growing angular velocity.

Here, it should be mentioned that in Ref. \cite{Braguta:2021jgn}, the normalization of the Polyakov-loop was not performed. In realistic calculations, a normalization factor needs to be introduced, resulting in a rescaling of the Polyakov-loop value, i.e, $\Phi_{nor}=C \cdot \Phi_{Latt}$. In our study, the introduction of a normalization factor has been shown through numerical calculations not to alter the qualitative behavior under investigation. For simplicity, we set the normalization factor $C=4$, which provides a more reliable value for the Polyakov-loop when the angular velocity is very small and the temperature is high. In the remainder of this article, $\Phi$ denotes the normalized Polyakov-loop.

Then, we construct a polarized Polyakov-loop potential ${\cal U}(\Phi ,\bar \Phi ,T,\omega)$  of the globally rotating pure gluon system,
which satisfies the $Z(3)$ center symmetry as in the pure gauge QCD Lagrangian. 
The normalized Polyakov-loop should be the solution of the gap equations
\begin{align}
    \frac{\partial {\cal U} (\Phi ,\bar{\Phi}, T, \omega)}{\partial\Phi}=\frac{\partial {\cal U} (\Phi ,\bar{\Phi}, T, \omega)}{\partial\bar\Phi}=0.
\end{align}
The expression of the potential is given as:
\begin{widetext}
\begin{eqnarray}\label{ppotential}
\frac{{\cal U}\left( {\Phi ,\bar{\Phi}, T, \omega} \right)}{T^4}=  -{C f\left( {T,\omega } \right)} {\left( {\frac{T}{{{T_0}}}} \right)^2}\Phi \bar \Phi  - \frac{1}{3}\left( {{\Phi ^3} + {{\bar \Phi }^3}} \right) + {C^{-1} f^{-1}\left( {T,\omega } \right)}{{{\left( {\frac{T}{{{T_0}}}} \right)}^{-2}}}{\Phi ^2}{{\bar \Phi }^2},
\end{eqnarray}
\end{widetext}
with the parameters listed in TABLE. \ref{table_parameters}.

\begin{table}
\centering
\begin{tabular}{c|p{7em} p{7.5em} p{7.5em}}
\hline
\hline
  $i$ &~~~~~~~~0 &~~~~~~~~~~~~2 &~~~~~~~~~~~~4   \\
\hline
  $a_{i}$ & 0.0454431 & -1.27942$\times10^{-5}$& -5.43339$\times10^{-9}$ \\
  $b_{i}$ & 46.8263 &  -0.0210165 &  -2.15394$\times10^{-5}$ \\
  $c_{i}$ & 1.00298 &  1.55157$\times10^{-4}$ & -5.99032$\times10^{-8}$ \\
  $d_{i}$ & 0.0600157 &  -5.74388$\times10^{-6}$  & -8.24192 \\
\hline
\hline
\end{tabular}
\vspace{0.4cm}
\caption{The parameters of the Polyakov-loop potential in Eq. (\ref{ppotential2}) and Eq. ( \ref{ppotential})  by fitting lattice data from Ref. \cite{Braguta:2021jgn}. }
\label{table_parameters}
\end{table}

\section{The 2-flavor PPNJL model\label{sec3}}

To explore the influence of rotation on QCD matter, we consider a rigid cylinder rotating around the $z$ axis with constant angular velocity $\omega$.  In this rotating frame of reference, the rotation can be characterized in relation to an external gravitational field.  The metric tensor can be utilized to describe the structure of space-time reads\\
\begin{eqnarray}
g_{\mu\nu}=\left(
\begin{array}{cccc}
 1-{\vec v}^{\, 2} & -v_1 & -v_2 & -v_3 \\
 -v_1 & -1 & 0 & 0 \\
 -v_2 & 0 & -1 & 0 \\
 -v_3 & 0 & 0 & -1 \\
\end{array}
\right),
\label{tensor}
\end{eqnarray}\\
where $v_i$ is the velocity and $v=\sqrt{v_1^2+v_2^2+v_3^2}$. In the context of a system with an angular velocity $\omega$ along the fixed $z$-axis,  we have $\vec{v}=\vec{\omega}\times \vec{x}$, thus $v_{3}=0$ in the metric tensor.

The Lagrangian in the two-flavor NJL model under rotation can be written as follows:
\begin{equation}
   {\cal L}_{\text{NJL}} =\\
    \sum_f\bar \psi_{f} \biggl[   {i\bar{\gamma} ^\mu (\partial _\mu+\Gamma_{\mu})- m+\gamma ^0 \mu } \biggr]\psi_{f} + G{\left( {\bar \psi \psi } \right)^2},
\end{equation}
here, $\psi$ is the quark field, $\bar{\gamma}^\mu=e_{a}^{\mu} \gamma^a$ with $e_{a}^{\mu}$ being the tetrads for spinors and $\gamma^{a}$ represents the gamma matrix, $\Gamma_\mu$ is defined as $\Gamma_\mu=\frac{1}{4}\times\frac{1}{2}[\gamma^a,\gamma^b] \ \Gamma_{ab\mu}$ which is the  spinor connection, where $\Gamma_{ab\mu}=\eta_{ac}(e^c_{\ \sigma} G^\sigma_{\mu\nu}e_b^{ \nu}-e_b^{\nu}\partial_\mu e^c_{ \nu})$, and $G^\sigma_{\mu\nu}$ is the affine connection determined by $g^{\mu\nu}$, $m$ is the bare quark mass matrix, $\mu$ denotes the chemical potential, and $G$ represents the coupling constants in the scalar channel. Here we only consider the scalar part, since we are mainly interested in the chiral phase transition.

The NJL model lacks gluon degrees of freedom. To effectively include the contribution from rotating gluodynamics, we add the NJL model on the rotating gluonic background given by Eq. (\ref{ppotential}). The Lagrangian of the PPNJL model takes the following form:

\begin{eqnarray}
{\cal L}_{\text{PPNJL}} = {{\cal L}_{\text{NJL}}} + \bar \psi {\gamma^{\mu} }{A_{\mu}}\psi  - {\cal U}(\Phi ,\bar \Phi ,T,\omega),
\label{covariantderivative}
\end{eqnarray}
here, we have included the coupling between the fermion fields and the gauge fields, and the effective polarized Polyakov-loop potential ${\cal U}(\Phi ,\bar \Phi ,T ,\omega )$, where the Polyakov-loop is a function of temperature and angular velocity given in Eq. (\ref{ppotential}), which effectively takes into account the roation effect on gluodynamics.  In this sense, the PPNJL model is different from the PNJL model, 

By selecting $e^{a}_{ \mu}=\delta^a_{\mu}+ \delta^a_{i}\delta^0_{\mu}  v_i$ and $e_{a}^{\mu}=\delta_a^{\mu} - \delta_a^{0}\delta_i^{\mu}  v_i$ (further details can be found in Refs. \cite{Fetter:2009zz,Jiang:2016woz}), the Lagrangian can be expanded to the first order of angular velocity, yielding the following expression:\\
\begin{widetext}
\begin{eqnarray}
{\cal L}_{\text{PPNJL}} = \bar \psi \left[ {i\gamma ^\mu  D _\mu   - m + \gamma ^0 \mu + \left( {\gamma ^0 } \right)^{ - 1} \left( {\left( {\mathord{\buildrel{\lower3pt\hbox{$\scriptscriptstyle\rightharpoonup$}}
\over \omega }  \times \mathord{\buildrel{\lower3pt\hbox{$\scriptscriptstyle\rightharpoonup$}}
\over x} } \right)\cdot\left( { - i\mathord{\buildrel{\lower3pt\hbox{$\scriptscriptstyle\rightharpoonup$}}
\over \partial } } \right) + \mathord{\buildrel{\lower3pt\hbox{$\scriptscriptstyle\rightharpoonup$}}
\over \omega } \cdot\mathord{\buildrel{\lower3pt\hbox{$\scriptscriptstyle\rightharpoonup$}}
\over S} _{4 \times 4} } \right)} \right]\psi + G {\left( {\bar \psi \psi } \right)^2 }- {\cal U}(\Phi, \bar \Phi, T, \omega)\label{Diracequation},
\end{eqnarray}
\end{widetext}
here the spin operator is given by $\mathord{\buildrel{\lower3pt\hbox{$\scriptscriptstyle\rightharpoonup$}} \over S} _{4 \times 4}  = \frac{1}{2}\left( {\begin{array}{*{20}c}
   {\mathord{\buildrel{\lower3pt\hbox{$\scriptscriptstyle\rightharpoonup$}} \over \sigma } } & 0  \\
   0 & {\mathord{\buildrel{\lower3pt\hbox{$\scriptscriptstyle\rightharpoonup$}} \over \sigma } }  \\
\end{array}} \right)$. From the equation above, it is evident that the free Dirac Lagrangian is modified due to the presence of the gluon field and rotation. The first term corresponds to the coupling between the quark field and gluon field, while the fourth term accounts for the orbital-rotation coupling effect and the spin-rotation coupling effect. The last term in the Lagrangian represents the effective Polyakov-loop potential. Here the covariant derivative is defined as  ${D_\mu } = {\partial _\mu } - i{A_\mu }$, with $A^{\mu}=\delta_{0}^{\mu}$, which determines the coupling between the Polyakov-loop and quarks. The Polyakov-loops $\Phi$ are obtained as follows,
\begin{eqnarray}
\Phi  = \frac{1}{{{N_c}}}\left\langle {{\rm{tr }}({\cal P} {\rm exp}[i {\int_0}^\beta
d\tau A_4{({\bar x},\tau)}])} \right\rangle ,
\end{eqnarray}
where $\beta=\frac {1}{T} $, $\cal P$ denotes path ordering, $A_4=iA_0$ is the temporal component of the Euclidean gauge field $(\bar A,A_4)$, and $ \bar \Phi = \Phi^{\dag}$. In this model, the quarks couple to a background (temporal) gauge field representing Polyakov-loop dynamics, and the Polyakov-loop serves as an order parameter for confinement. When performing the mean field approximation and employing the technique of path integral formulation for Grassmann variables theory, the grand thermodynamic potential reads, 
\begin{widetext}
\begin{eqnarray}
\label{eq_Omega}
\Omega_{\text{PPNJL}} & = & G\langle \bar{q}q\rangle^2-\frac{3}{2\pi^2}\sum_{n=-\infty}^{\infty}\int_{0}^{\Lambda}p_tdp_t\int_{-\sqrt{\Lambda^2-p_t^2}}^{\sqrt{\Lambda^2-p_t^2}}dp_z \biggl(J_{n+1}(p_tr)^2+J_{n}(p_tr)^2\biggr)\epsilon_n \nonumber\\
&-&\frac{T}{2\pi^2}\sum_{n=-\infty}^{\infty}\int p_tdp_t dp_z  \Biggl[ \biggl(J_{n+1}(p_tr)^2+J_{n}(p_tr)^2\biggr) \text{log}\biggl(1+3\Phi \text{exp}(-\frac{\epsilon_n-\mu}{T})+3\bar{\Phi}\text{exp}(-2\frac{\epsilon_n-\mu}{T})+\text{exp}(-3\frac{\epsilon_n-\mu}{T})\biggr) \Biggr]\nonumber\\
&-&\frac{T}{2\pi^2}\sum_{n=-\infty}^{\infty}\int p_tdp_t dp_z  \Biggl[ \biggl(J_{n+1}(p_tr)^2+J_{n}(p_tr)^2\biggr)  \text{log}\biggl(1+3\bar{\Phi} \text{exp}(-\frac{\epsilon_n+\mu}{T})+3\Phi\text{exp}(-2\frac{\epsilon_n+\mu}{T})+\text{exp}(-3\frac{\epsilon_n+\mu}{T})\biggr) \Biggr]\nonumber\\
&+& {\cal U}(\Phi ,\bar \Phi ,T,\omega ). 
\end{eqnarray}
\end{widetext}

Here, for simplicity, we have introduced the quark quasiparticle energy
${\varepsilon _n} = \sqrt {M^2 + p_t^2 + p_z^2}  - \left( {\frac{1}{2} + n} \right)\omega$ with the dynamic  quark mass ${M} = {m} - 2G\left\langle {\bar qq} \right\rangle $. From the grand thermodynamic potential, the equations of motion for the mean fields $\left\langle\bar qq \right\rangle $, $\Phi$, and $\bar \Phi$ can be derived as

\begin{eqnarray}
\frac{{\partial \Omega_{\text{PPNJL}} }}{{\partial \left\langle {\bar qq} \right\rangle }} = 0,~~~ \frac{{\partial \Omega_{\text{PPNJL}} }}{{\partial \Phi }} = 0,~~~ \frac{{\partial \Omega_{\text{PPNJL}} }}{{\partial \bar \Phi }} = 0.
\end{eqnarray}
These coupled equations are then solved as functions of temperature $T$, quark chemical potential $\mu$, and angular velocity $\omega$. The detailed expressions for these stationary condition equations are provided below: \\
\\

\begin{widetext}
\begin{eqnarray}
0 &=&2G\langle \bar{q}q\rangle+\frac{3}{2\pi^2}\sum_{n=-\infty}^{\infty}\int_{0}^{\Lambda}p_tdp_t\int_{-\sqrt{\Lambda^2-p_t^2}}^{\sqrt{\Lambda^2-p_t^2}}dp_z \biggl(J_{n+1}(p_tr)^2+J_{n}(p_tr)^2\biggr)\left( {  \frac{{2G M}}{{\sqrt {p_t^2 + p_z^2 + M^2} }}} \right)\nonumber\\
&-&\frac{1}{2\pi^2}\sum_{n=-\infty}^{\infty}\int dp_tdp_z p_t \biggl(J_{n+1}(p_tr)^2+J_{n}(p_tr)^2\biggr)\left(   \frac{6G M}{\sqrt {p_t^2 + p_z^2 + M^2}} \right)\nonumber\\
\nonumber\\
&\times&\left(\frac{ \text{exp}(-3\frac{\epsilon_n-\mu}{T}) + \Phi \text{exp}(-\frac{\epsilon_n-\mu}{T})  + 2
\bar{\Phi}\text{exp}(-2\frac{\epsilon_n-\mu}{T}) }
{1+\text{exp}(-3\frac{\epsilon_n-\mu}{T})+3\Phi \text{exp}(-\frac{\epsilon_n-\mu}{T})+3\bar{\Phi}\text{exp}(-2\frac{\epsilon_n-\mu}{T})}+     \frac{ \text{exp}(-3\frac{ \varepsilon _n+\mu  }{T}) + 2\Phi{\text{exp}(-2\frac{\varepsilon _n+\mu}{T})}  + \bar \Phi
{\text{exp}(-\frac{{ \varepsilon _n+\mu }}{T})} }
{1 + \text{exp}(-3\frac{ \varepsilon _n+\mu}{T}) + 3\Phi{\text{exp}(-2\frac{\varepsilon _n+\mu}{T})}  + 3\bar \Phi{\text{exp}(-\frac{\varepsilon _n+\mu}{T})} }\right),\nonumber\\
\end{eqnarray}\\

\begin{eqnarray}
0 &=& - \frac{{1}}{{2{\pi ^2}}}T\sum_{n=-\infty}^{\infty}\int dp_tdp_z p_t \biggl(J_{n+1}(p_tr)^2+J_{n}(p_tr)^2\biggr)  \nonumber \\
&\times&\left( {\frac{3{\text{exp}(-\frac{{\varepsilon _n}-{\mu}}{T})}}{1+\text{exp}(-3\frac{\epsilon_n-\mu}{T})+3\Phi \text{exp}(-\frac{\epsilon_n-\mu}{T})+3\bar{\Phi}\text{exp}(-2\frac{\epsilon_n-\mu}{T})} +        \frac{{3\text{exp}(-2\frac{\varepsilon _n+\mu}{T})}}{{1 + \text{exp}(-3\frac{ \varepsilon _n+\mu}{T}) + 3\Phi{\text{exp}(-2\frac{\varepsilon _n+\mu}{T})}  + 3\bar \Phi{\text{exp}(-\frac{\varepsilon _n+\mu}{T})} }}} \right) \nonumber \\
\nonumber \\
&+& T^4 \left(-{C f\left( {T,\omega } \right)} {\left( {\frac{T}{{{T_0}}}} \right)^2}{\bar \Phi }  - \Phi^{2} + 2{C^{-1} f^{-1}\left( {T,\omega } \right)}{{{\left( {\frac{T}{{{T_0}}}} \right)}^{-2}}}{\Phi }{{\bar \Phi }^2}\right),
\end{eqnarray}\\
\\

\begin{eqnarray}
0 &=& - \frac{{1}}{{2{\pi ^2}}}T\sum_{n=-\infty}^{\infty}\int dp_tdp_z p_t \biggl(J_{n+1}(p_tr)^2+J_{n}(p_tr)^2\biggr)  \nonumber \\
&\times&\left( \frac{3\text{exp}(-2\frac{{\varepsilon _n}-{\mu}}{T})}{1+\text{exp}(-3\frac{\epsilon_n-\mu}{T})+3\Phi \text{exp}(-\frac{\epsilon_n-\mu}{T})+3\bar{\Phi}\text{exp}(-2\frac{\epsilon_n-\mu}{T})} +        \frac{{3\text{exp}(-\frac{\varepsilon _n+\mu}{T})}}{{1 + \text{exp}(-3\frac{\varepsilon _n+\mu}{T})}  + 3\Phi\text{exp}(-2\frac{\varepsilon_n+\mu}{T})  + 3\bar{\Phi}\text{exp}(-\frac{\varepsilon_n+\mu}{T})} \right) \nonumber \\
\nonumber \\
&+& T^4 \left(-{Cf(T,\omega)} \frac{T^2}{T_0^2}\Phi  - \bar{\Phi}^2+ 2{C^{-1} f^{-1}\left( {T,\omega } \right)} \Phi ^2\bar \Phi\right).
\end{eqnarray}
\end{widetext}

\section{Spin alignment of vector meson $\rho$  \label{sec4}}

In the following section, we study on the spin alignment of the vector meson $\rho$ under rotation. We will briefly discuss the phenomenon of vector meson spin alignment under rotation. The spin states of vector mesons are described by the $3 \times 3$ spin-density matrix $\rho_{ij}$. However, the polarization of vector mesons cannot be directly measured, due to the parity-conserving of their strong decay. It is challenging to ascertain whether the meson spin aligns parallel or anti-parallel to the angular momentum, and with only experimental access to $\rho_{00}$. 

Experimentally, the spin alignment of vector mesons can be measured by the angular distribution of decay products in the vector meson decaying to two spinless particles  \cite{Schilling:1969um}:
\begin{eqnarray}
\frac{{{\rm{d}}N}}{{{\rm{d}}\cos {\theta ^*}}} = \frac{3}{4}[1 - {\rho _{00}} + \left( {3{\rho _{00}} - 1} \right){\cos ^2}{\theta ^*}],
\end{eqnarray}
where $\theta ^*$ is the polar angle between the quantization axis and the momentum direction of the decay particle. 
In a simple recombination picture, the $\rho$ vector meson is produced by the simple coalescence of a quark and antiquark with polarizations $P_{q}$ and $P_{\bar{q}}$ respectively. The meson spin alignment is given by \cite{Liang:2004xn}:
\begin{eqnarray}
{\rho _{00}} = \frac{{1 - {P_q}{P_{\bar q}}}}{{3 + {P_q}{P_{\bar q}}}},
\end{eqnarray}
and when ${P_q}{P_{\bar q}}$ is small, the spin alignment of $\rho$ can be approximated as:
\begin{equation}
\rho _{00}   \approx \frac{1}{3} - \frac{4}{9}P_q P_{\bar q},
\end{equation}
where $P_q$ is the spin polarization of quark, and we simply take the following definition:
\begin{equation}
P_q  = \frac{{N_ \uparrow ^ +   - N_ \downarrow ^ +  }}{{N_ \uparrow ^ +   + N_ \downarrow ^ +  }},
\end{equation}
here, $N^{+/-}_{\uparrow/\downarrow}$ denote the quark/anti-quark number density with spin up/down, respectively, which can be extracted by taking the partial derivative of $ \Omega_{\text{PPNJL}}$ Eq. (\ref{eq_Omega}) with respect to $\mu$. The detailed expressions for $N_ \uparrow ^ + ,N_ \downarrow ^ + ,N_ \uparrow ^ - ,N_ \downarrow ^ - $ are listed as follows:

\begin{widetext}
\begin{eqnarray}
N_ \uparrow ^ + & = & \frac{1}{2\pi^2}\sum_{n=-\infty}^{\infty}\int dp_tdp_z p_t J_{n}(p_tr)^2 \frac{3\Phi \text{exp}(-\frac{\epsilon_n-\mu}{T})+6\bar{\Phi}\text{exp}(-2\frac{\epsilon_n-\mu}{T})+3\text{exp}(-3\frac{\epsilon_n-\mu}{T})}{1+3\Phi \text{exp}(-\frac{\epsilon_n-\mu}{T})+3\bar{\Phi}\text{exp}(-2\frac{\epsilon_n-\mu}{T})+\text{exp}(-3\frac{\epsilon_n-\mu}{T})},\\
\nonumber\\
N_ \downarrow ^ + & = & \frac{1}{2\pi^2}\sum_{n=-\infty}^{\infty}\int dp_tdp_z p_t J_{n+1}(p_tr)^2 \frac{3\Phi \text{exp}(-\frac{\epsilon_n-\mu}{T})+6\bar{\Phi}\text{exp}(-2\frac{\epsilon_n-\mu}{T})+3\text{exp}(-3\frac{\epsilon_n-\mu}{T})}{1+3\Phi \text{exp}(-\frac{\epsilon_n-\mu}{T})+3\bar{\Phi}\text{exp}(-2\frac{\epsilon_n-\mu}{T})+\text{exp}(-3\frac{\epsilon_n-\mu}{T})},\\
\nonumber\\
N_ \uparrow ^ - & = & \frac{1}{2\pi^2}\sum_{n=-\infty}^{\infty}\int dp_tdp_z p_t J_{n}(p_tr)^2 \frac{-3\bar{\Phi} \text{exp}(-\frac{\epsilon_n+\mu}{T})-6\Phi\text{exp}(-2\frac{\epsilon_n+\mu}{T})-3\text{exp}(-3\frac{\epsilon_n+\mu}{T})}{1+3\bar{\Phi} \text{exp}(-\frac{\epsilon_n+\mu}{T})+3\Phi\text{exp}(-2\frac{\epsilon_n+\mu}{T})+\text{exp}(-3\frac{\epsilon_n+\mu}{T})},\\
\nonumber\\
N_ \downarrow ^ - & = & \frac{1}{2\pi^2}\sum_{n=-\infty}^{\infty}\int dp_tdp_z p_t J_{n+1}(p_tr)^2 \frac{-3\bar{\Phi} \text{exp}(-\frac{\epsilon_n+\mu}{T})-6\Phi\text{exp}(-2\frac{\epsilon_n+\mu}{T})-3\text{exp}(-3\frac{\epsilon_n+\mu}{T})}{1+3\bar{\Phi} \text{exp}(-\frac{\epsilon_n+\mu}{T})+3\Phi\text{exp}(-2\frac{\epsilon_n+\mu}{T})+\text{exp}(-3\frac{\epsilon_n+\mu}{T})}.
\end{eqnarray}
\end{widetext}

Then, the spin alignment of $\rho$ can be approximately expressed as:
\begin{eqnarray}
\rho _{00}   = \frac{1}{3} - \frac{4}{9}\frac{{N_ \uparrow ^ +   - N_ \downarrow ^ +  }}{{N_ \uparrow ^ +   + N_ \downarrow ^ +  }}\frac{{N_ \uparrow ^ -   - N_ \downarrow ^ -  }}{{N_ \uparrow ^ -   + N_ \downarrow ^ -  }}.
\end{eqnarray}

\section{Numerical results and discussions \label{sec5}}

\begin{figure}
\subfigure[]{\includegraphics[width=0.5\textwidth]{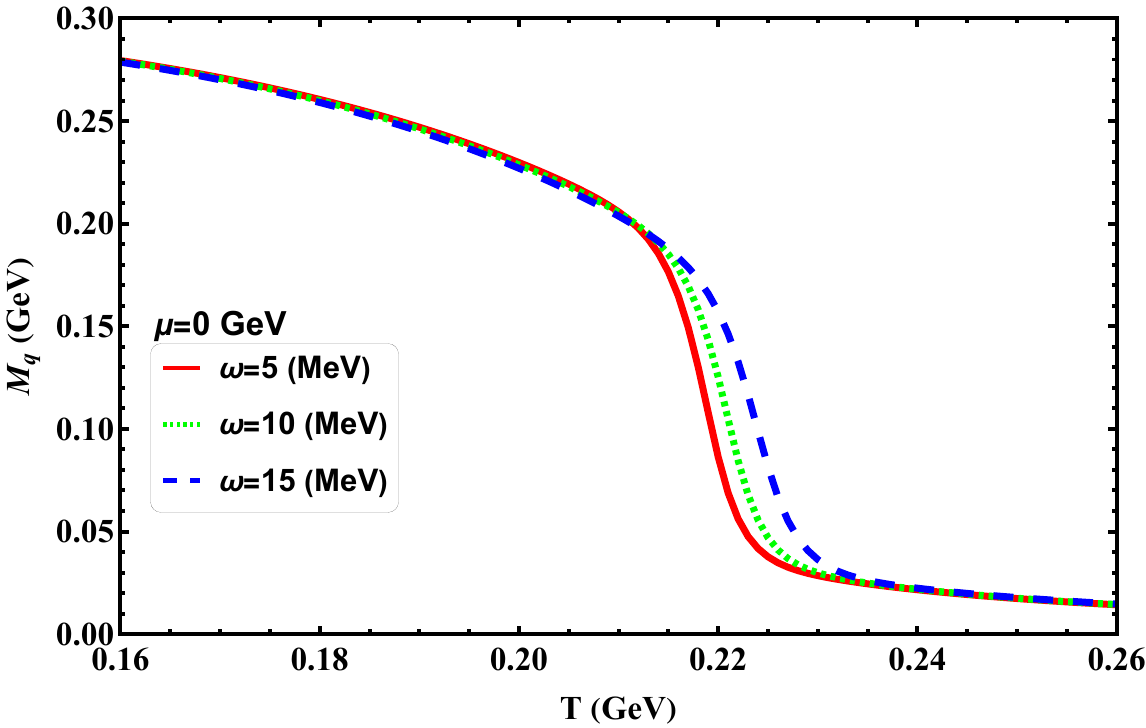}}
\vspace{0.8cm}
\subfigure[]{\includegraphics[width=0.5\textwidth]{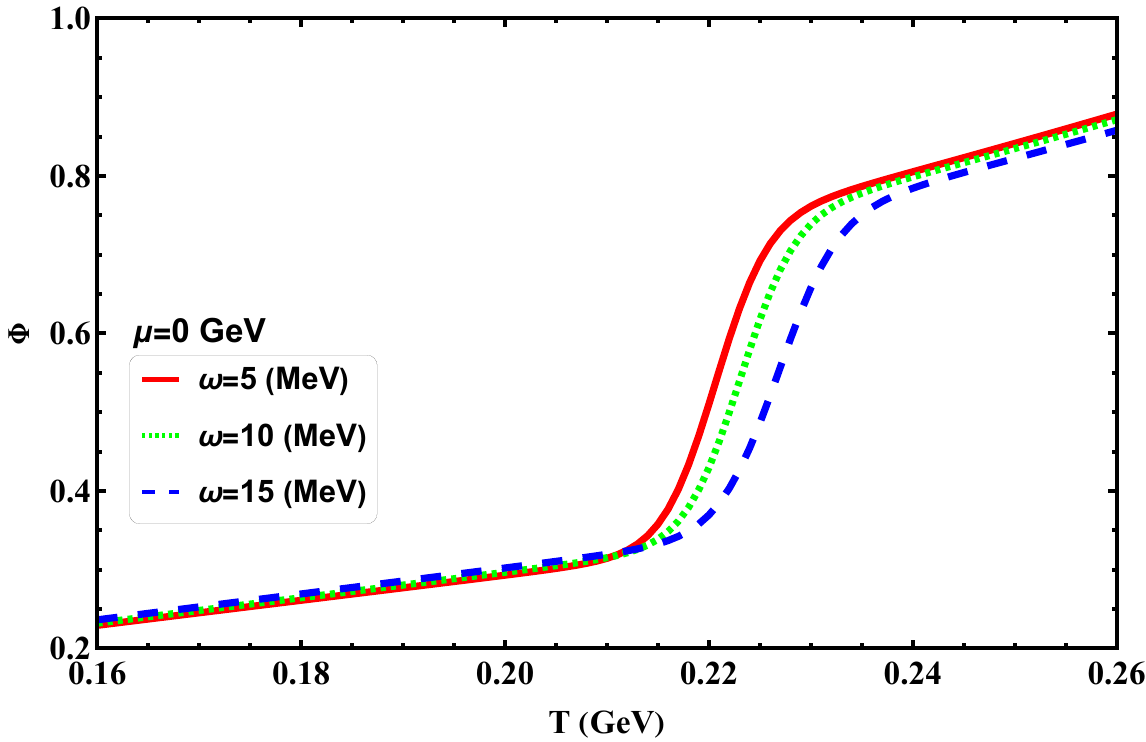}}
\vspace{0.8cm}
\subfigure[]{\includegraphics[width=0.5\textwidth]{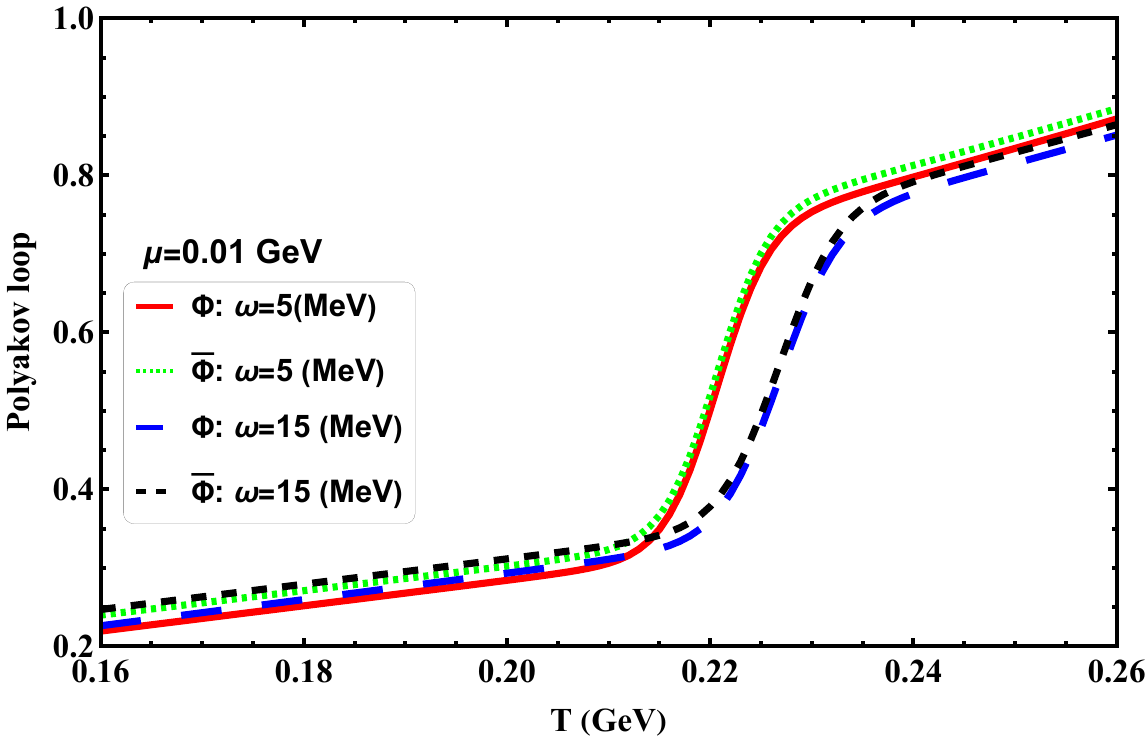}}
\caption[]{(Color online) (a)The light quark effective mass and (b)Polyakov-loop as functions of temperature $T$ at zero quark chemical potential $\mu=0$ GeV for different angular velocities, $\omega=5, 10, 15$ MeV; (c)Polyakov-loop as a function of temperature $T$ at nonzero quark chemical potential $\mu=0.01$ GeV for different angular velocities, $\omega=5,  15$ MeV.}
\label{MPhitoT.pdf}
\end{figure}

In this section, we present the numerical results obtained from the PPNJL model under rotation. For the Fermionic sector, we have selected the parameters $m=0.005$ GeV, $\Lambda=0.65$ GeV, and $G=4.93~ \text{GeV}^{-2}$, as reported in Ref. \cite{Klevansky:1992qe}, to match physical observations. Additionally, we denote the $z$-angular-momentum quantum number as $n=0,\pm1,\pm2...$. In principle, the sum over all values of $n$ is necessary; however, the rapid convergence of these expressions allows us to limit the sum over $n$ from $-5$ to $5$. It is important to note that in the PPNJL model, the value of $T_{0}$ can be rescaled over a wide range, and here we set $T_{0}=0.22$ GeV. The radius is set to $r=0.1$ GeV$^{-1}$, ensuring that $\omega r<1$ in all calculations.

Fig. \ref{MPhitoT.pdf} (a) and (b) depict the light quark effective mass and Polyakov-loop as functions of temperature $T$ at quark chemical potential $\mu=0$ GeV for different angular velocities, respectively. The quark condensate is related to the spontaneous breaking of chiral symmetry, and in the 2-flavor PPNJL model, there exists a simple relationship between the condensate and the quark effective mass, given by ${M} = {m} - 2G\left\langle {\bar qq} \right\rangle$. Meanwhile, the Polyakov-loop is associated with confinement-deconfinement. Within the temperature range considered, we observe that at low temperatures, both the quark effective mass and the Polyakov-loop exhibit weak dependence on temperature for the different angular velocities selected. As the temperature continues to increase and exceeds the transition region, the light quark effective mass tends to approach its current mass, and the value of Polyakov-loop tends to approach $1$. This indicates the restoration of chiral symmetry and the gradual recovery of deconfinement at high temperatures. Surprisingly, when considering the influence of the gluonic rotational contribution, we observe that the effective quark mass is enhanced by the angular velocity in the chiral phase transition region, while the Polyakov-loop is suppressed by the angular velocity in the confinement-deconfinement phase transition region. It is worth mentioning that in the general NJL model or in the PNJL model under rotation, where the gluonic rotational contribution is not taken into account, rotation tends to suppress the quark condensate. In contrast to the majority of effective model predictions, when considering the  polarization of the gluodynamics induced by the rotation, our results in the PPNJL model indicate an opposite behavior. In the transition region, there is an increase in the chiral condensate and a decrease in the Polyakov-loop with rotation. This suggests that rotation acts as a catalyst for chiral symmetry breaking and confinement. Next, we turn to the situation of non-zero chemical potential, Fig. \ref{MPhitoT.pdf} (c) shows the Polyakov-loop as a function of temperature $T$ at nonzero quark chemical potential $\mu=0.01$ GeV for different angular velocities, $\omega=5, 15$ MeV. When we introduce a non-zero chemical potential, we observe that $\Phi$ and $\bar \Phi$ differ from each other and for a fixed angular velocity $\bar \Phi>\Phi$.

\begin{figure}
{
\includegraphics[width=0.5\textwidth]{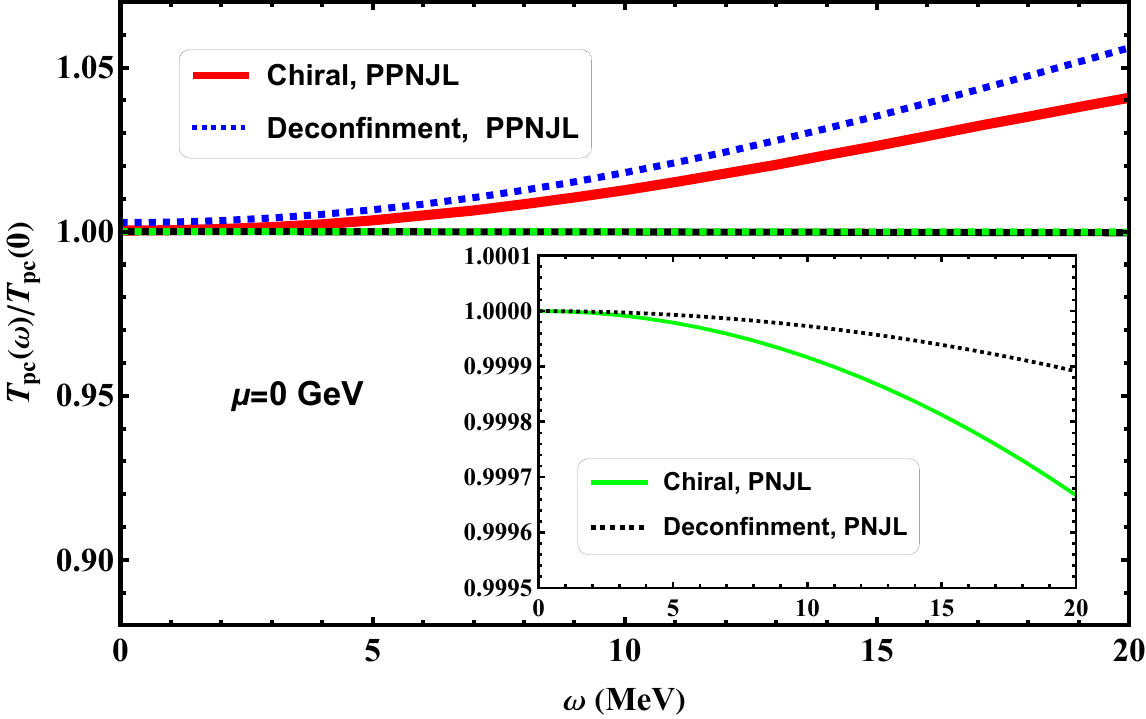}
}
\caption[]{(Color online) The scaled pseudocritical temperatures for chiral and deconfinement phase transition , determined by the quark condensate and Polyakov-loop, as functions of the angular velocity at zero chemical potential, and  both the Polyakov-loop potential obtained with and without explicit rotational dependence are considered, which corresponding to the PPNJL and PNJL models under rotation, respectively.}
\label{TC.pdf}
\end{figure}

\begin{figure}
{
\includegraphics[width=0.45\textwidth]{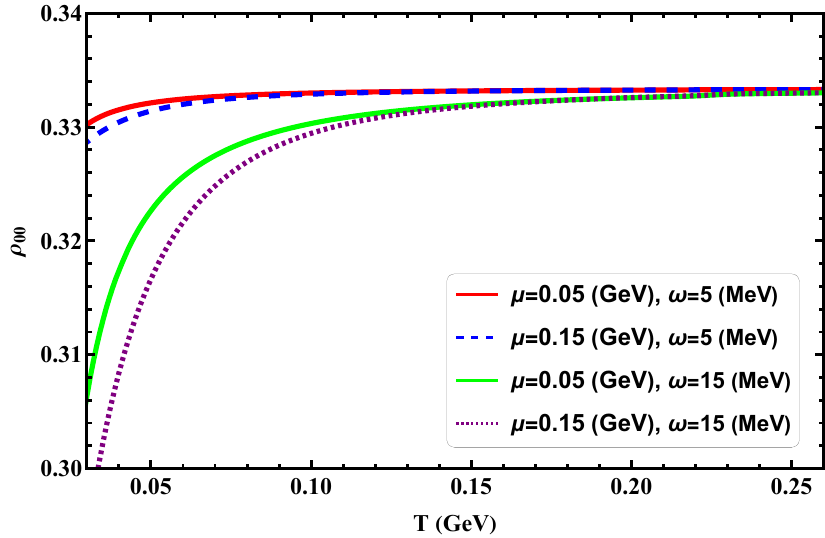}
}
\caption[]{(Color online) The $\rho$ meson spin alignment as a function of temperature $T$ at $\mu=0.05$ GeV,  and $\mu=0.15$ GeV for $\omega=5, 15$ MeV, respectively.}
\label{alignment.pdf}
\end{figure}

The impact of the rotational contribution of gluons on the phase transition is illustrated in Fig. \ref{TC.pdf}, which shows the scaled pseudocritical temperatures of the  chiral phase transition and deconfinement phase transition as a function of the angular velocity at zero chemical potential. The figure considers the cases of Polyakov-loop potential with and without explicit rotational dependence, denoted as ${\cal U} = {\cal U}(\Phi ,\bar \Phi ,T,\omega)$ and ${\cal U} = {\cal U}(\Phi ,\bar \Phi ,T)$, respectively (here the exact expression for $\mathcal{U}(\Phi, \bar \Phi, T)$ is taken from Ref. \cite{Ratti:2005jh}). It is evident that with the explicit rotational dependence, the pseudocritical temperatures of both the chiral phase transition and confinement-deconfinement phase transition increase with the angular velocity, which is consistent with lattice QCD simulations results \cite{Braguta:2022str,Yang:2023vsw}. When considering the Polyakov-loop potential without explicit rotational dependence, the dashed line and dotted line almost overlap, to further illustrate the rotation dependence of pseudo-critical temperatures in this case, an enlarged sub-figure is included in Fig. \ref{TC.pdf}. It can be clearly seen from the sub-figure that the pseudocritical temperatures of both chiral and deconfinement phase transitions decrease with increasing angular velocity. The chiral phase transition has been extensively studied in the NJL model and holographic QCD models, and it has been observed that the critical temperature of chiral phase transition decreases as the angular velocity increases. However, when considering the contribution from the polarized gluodynamics under rotation, it has the opposite effect on the critical temperatures. The opposite effect indicates that the rotating gluons has important effect on rotating quarks, and it is necessary to take into account the contribution of rotating gluons to fully understand the characteristics of a rotating QCD medium.

Fig. \ref{alignment.pdf} depicts the influence of rotating gluons on the spin alignment of the $\rho$ meson as a function of temperature $T$ at quark chemical potentials $\mu=0.05$ GeV and $\mu=0.15$ GeV for different angular velocities. Any deviation of the spin alignment from $1/3$ indicates the polarization of the vector mesons along a particular direction. It is evident that at high temperatures, the spin alignment approaches $1/3$, while at low temperatures, rotation can significantly enhance the deviation from $1/3$. When comparing the results corresponding to different quark chemical potentials at a fixed angular velocity, a larger quark chemical potential can lead to a larger deviation from $1/3$. However, compared to the
contribution from the angular velocity (note that here, the value of $\omega$ is in MeV, while $\mu$ is in GeV), the deviation induced by the chemical potential is less obvious. Our results indicate that in the 2-flavor PPNJL model under rotation, the rotating gluon may play a 
significant role in the spin alignment of the $\rho$ meson.

\begin{figure}
{
\includegraphics[width=0.5\textwidth]{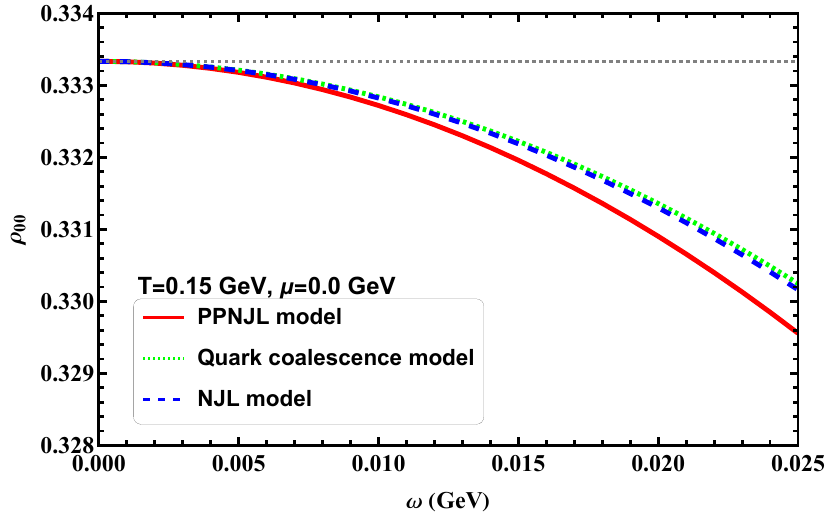}
}
\caption[]{(Color online) The $\rho$ meson spin alignment as a function of angular velocity $\omega$ at  $T=0.15$ GeV and  $\mu=0.0$ GeV in the PPNJL model and comparing with the NJL model and quark coalescence model. The gray dashed line is $\rho_{00}=\frac{1}{3}$.}
\label{allignmentinthreemodels.pdf}
\end{figure}

From \cite{Wei:2023pdf}, it has been summarized that the spin alignment of vector meson (for both $\rho$ and $\phi$) in the NJL model at $T=150~ {\rm MeV}$ has the relation of
\begin{equation}
\rho_{00}(\omega)=\frac{1}{3}-5.10 \omega^2+39.62 \omega^4,
\end{equation} and in the quark coalescence model \cite{Yang:2017sdk} the spin alignment of vector meson has the form of
\begin{equation}
 \rho_{00}(\omega)=\frac{1}{3}-\frac{1}{9}(\beta \omega)^2,
 \end{equation}
 with $\beta=1/T$. In Fig. \ref{allignmentinthreemodels.pdf} we show the spin alignment of $\rho$ meson as a function of $\omega$ in the PPNJL model, NJL model and quark coalescence model. It is noticed that the spin alignment of vector meson has the negative deviation of $\rho_{00}-1/3$ under rotation, in the case of considering the effect of rotating gluons, the spin alignment of vector meson is more significantly deviated from $1/3$.

\section{CONCLUSIONS\label{sec6}}

 In this work, in order to effectively take into account the rotating gluon background, by using the extrapolation method, we construct the polarized Polyakov-loop potential induced by rotation based on the lattice results at finite imaginary angular velocity, and add the 2-flavor NJL model on the rotating gluonic background. We explore the QCD phase transitions and spin alignment of vector mesons using this 2-flavor Polarized-Polyakov-loop Nambu-Jona-Lasinio (PPNJL) model under rotation. It is found that the critical temperatures for both chiral phase transition and deconfinement phase transition increase with the angular velocity, which is  consistent with lattice results, and opposite to results in the NJL/PNJL model and holographic QCD model. Our findings reveal that the quark condensate and critical temperature of the phase transition are sensitive to the effects of rotating gluons.

 Furthermore, we calculate the pin alignment of vector $\rho$ meson, and find that the spin alignment of vector meson has a negative deviation of $\rho_{00}-1/3$ under rotation,  and the deviation in the PPNJL model is much more significant than that in the NJL model and the quark coalescence model, indicating the significant contribution from the rotating gluons.

Our study suggests that the contribution of rotating gluons is essential for correctly understanding the properties of rotating QCD. The effect of rotating gluons is also anticipated to make significant contributions to various theoretical investigations. Recent studies in Refs. \cite{Jiang:2021izj,Jiang:2023hdr} have demonstrated that in the NJL model, the influence of rotating gluons can be effectively taken into account by the running of the effective coupling, resulting in a modification of the critical temperature behavior in the rotating system. Alternative simulation method, e.g., Ref. \cite{Cao:2023olg} discussed the reliability of the analytic continuation of the phase diagram from imaginary rotation to real rotation. These investigations would help determine the Polyakov-loop potential and unravel other related puzzles, thus greatly enhancing our understanding of the properties of rotating QCD. We hope that our study will stimulate further theoretical and lattice QCD simulation studies in the future.

\section*{Acknowledgements}
We would like to thank Xu-Guang Huang, Ji-Chong Yang and Anping Huang for useful discussions. The work has been supported by the National Natural Science Foundation of China (NSFC) with Grant No.s: 12235016, 12221005 and the Strategic Priority Research Program of Chinese Academy of Sciences under Grant No XDB34030000.

\bibliography{ref-lib}
\end{document}